# Influence of Humidity on the Resistive Switching of Hexagonal Boron Nitride-Based Memristors

*Lukas Völkel[a], Rana Walied Ahmad[b], Alana Bestaeva[b], Dennis Braun[a], Sofia Cruces[a], Jimin Lee[a], Sergej Pasko[c], Simonas Krotkus[c], Michael Heuken[c], Stephan Menzel[b], Max C. Lemme[a,d]\**

[a] Chair of Electronic Devices, RWTH Aachen University, Otto-Blumenthal-Str. 25, 52074 Aachen, Germany.

[b] Jülich Aachen Research Alliance (JARA)-Fit and Peter Grünberg Institute (PGI-7), Forschungszentrum Jülich GmbH, Jülich, Germany

[c] AIXTRON SE, Dornkaulstraße 2, 52134 Herzogenrath, Germany

[d] AMO GmbH, Advanced Microelectronic Center Aachen, Otto-Blumenthal-Str. 25, 52074 Aachen, Germany.

\*Email: max.lemme@eld.rwth-aachen.de

**Abstract:**

Two-dimensional material-based memristors have recently gained attention as components of future neuromorphic computing concepts. However, their surrounding atmosphere can influence their behavior. In this work, we investigate the resistive switching behavior of hexagonal boron nitride-based memristors with active nickel electrodes under vacuum conditions. Our cells exhibit repeatable, bipolar, nonvolatile switching under voltage stress after initial forming, with a switching window $> 10^3$ under ambient conditions. However, in a vacuum, the forming is suppressed, and hence, no switching is observed. Compact model simulations can reproduce the set kinetics of our cells under ambient conditions and predict highly suppressed resistive switching in a water-deficient environment, supporting the experimental results. Our findings have important implications for the application of h-BN-based memristors with electrochemically active electrodes since semiconductor chips are typically processed under high vacuum conditions and encapsulated to protect them from atmospheric influences.

### Introduction

Two-dimensional (2D) materials have attracted attention as the active medium in resistive switching devices [1–3]. Hexagonal boron nitride (h-BN) is of special interest, as it is an atomically thin insulator with a band gap of approximately 6 eV [4], promising low energy consumption [5], fast switching [6] and a large switching window [7]. h-BN has been shown to exhibit both nonvolatile and volatile (threshold) resistive switching behavior [7]. The predominant resistive switching mechanism discussed in the literature is metallic filament formation under voltage stress [8–10], analogous to electrochemical metallization (ECM) cells based on more mature metal oxides [11]. Ambient air molecules such as water can significantly influence the resistive switching behavior of metal oxide-based ECM cells [12]. Furthermore, 2D materials are known to be affected by environmental contaminants [13,14], and Wang *et al.* reported the aging of h-BN memristors in humid atmospheres [15]. However, the effect of vacuum on the resistive switching behavior of h-BN memristors has rarely been explored. Here, we investigate the influence of ambient conditions on nonvolatile h-BN memristors by comparing their resistive switching behavior under ambient air and vacuum environments. We support our experimental findings by simulating the structures in ambient and vacuum-like environments with a JART ECM compact model [16,17].

**Results and Discussion**

For this study, we fabricated vertical memristors with h-BN sandwiched between two metal electrodes. A schematic view of the material stack is depicted in Fig. 1a. The memristors consists of a 50 nm titan/palladium (Ti/Pd) bottom electrode (BE), 3 nm h-BN, and a 60 nm nickel/aluminum (Ni/Al) top electrode (TE) on a silicon (Si) chip with 300 nm thermally oxidized silicon dioxide ($SiO_2$). More fabrication details can be found in the experimental section. A top-view optical micrograph of a fabricated cross-point device is shown in Fig. 1b. The active area of the devices was ~30 µm² (Fig. 1b). The thickness of the transferred h-BN was measured via atomic force microscopy (AFM) to be ~3 nm (Fig. 1c). Raman measurements were conducted on top of the transferred film and on the substrate to confirm the presence of h-BN (Fig. 1d). The Raman spectrum exhibits the characteristic h-BN $E_{2g}$ Raman peak at 1368 cm$^{-1}$ [18], which is absent when it is measured next to the h-BN film on the same chips.

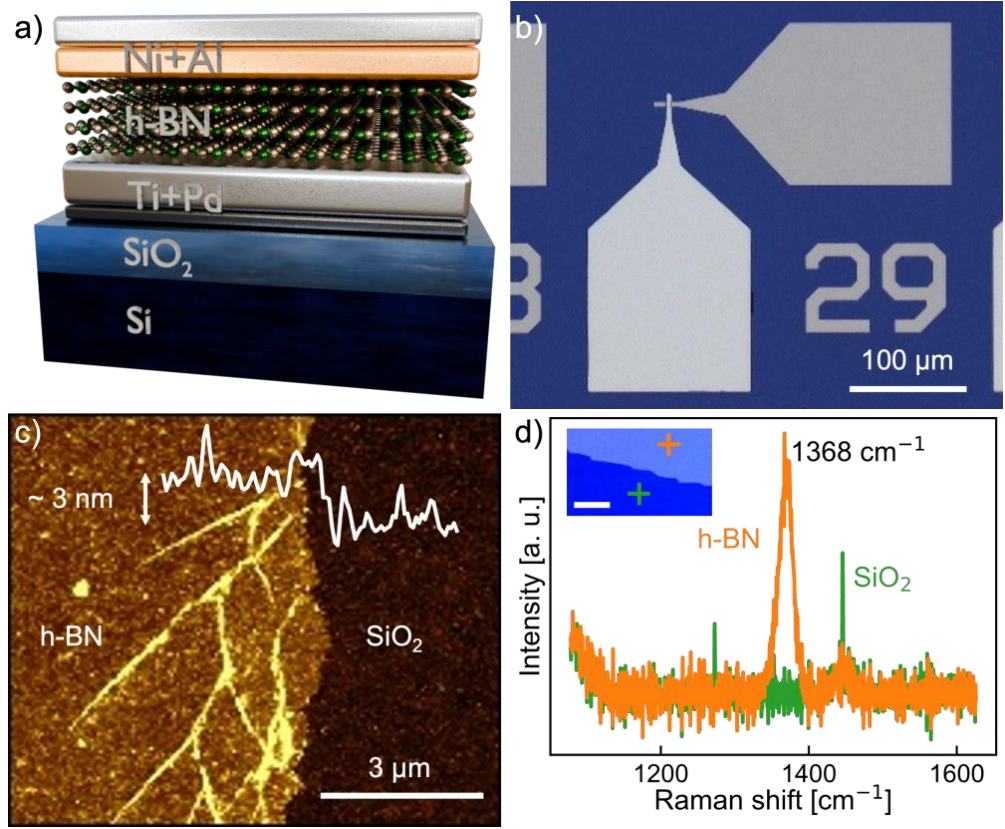

**Figure 1:** Device fabrication. a) Schematic of our fabricated Pd/h-BN/Ni memristors on a Si/SiO$_2$ substrate. b) Micrograph of a crosspoint device. The h-BN covers the entire area below the top electrode. c) AFM measurement of h-BN transferred to SiO$_2$. The white line at the top shows the thickness profile along the edge. The h-BN thickness is ~3 nm. d) Comparative Raman measurements of h-BN and SiO$_2$. The measurement positions are marked in the micrograph (inset). The scale bar is 10 μm. The Raman spectrum of the h-BN shows a characteristic peak at 1368 cm$^{-1}$.

We performed electrical measurements in a cryogenic probe station "CRX-6.5K" from LakeShore Cryotronics connected to a Tektronix "4200A-SCS" semiconductor parameter analyzer. The voltage was applied to the TE while the BE was grounded. Initially, our devices were in a high resistance state (HRS). We performed DC current-voltage (*I-V*) measurements by sweeping the voltage from 0 V to 2.5 V and back with a sweep rate of ~ 230 mV/s, then to -1.4 V and back with a sweep rate of ~ 100 mV/s, while measuring the resulting device current. The voltage stress led to repeatable bipolar resistive switching with a set voltage of ~ 1.7 V and a reset voltage of ~ -0.8 V. Fig. 2 shows 500 successively measured DC *I-V* sweeps. The inset shows the low resistance state (LRS) and HRS values over the cycle number with an average resistance window of > $10^3$.

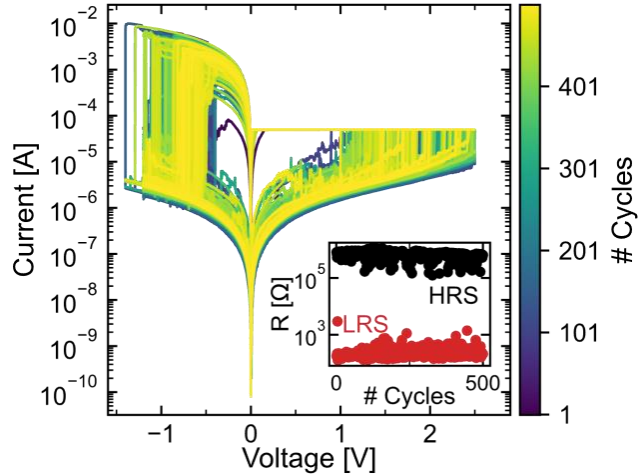

**Figure 2:** Five hundred successively measured *I-V* curves showing repeatable bipolar resistive switching of one memristor. The inset plot shows the resistances of the high (HRS) and low resistive states (LRS) with an increasing number of cycles.

We performed DC *I-V* measurements under vacuum to investigate the influence of ambient air on the resistive switching behavior. The first switching event observed in a pristine memristor under voltage stress is typically called "forming". This term arises from the formation of the first filament in filamentary switching devices, which establishes the foundation for subsequent resistive switching cycles [19]. Fig. 3 shows the forming attempts (*I-V* sweeps) of eight devices under both ambient (left) and vacuum (right) conditions. Vacuum conditions were achieved by pumping down the measurement chamber for 67 h. The pressure during the vacuum measurements was 4.8 x $10^{-5}$ mbar. The switching parameters forming voltage $V_{form}$ and current $I_{form}$ that lead to the successful forming of the devices in ambient air are shaded in gray in both plots. Under ambient conditions, the initial filament was formed for all devices when the applied voltage and the device current exceeded +2.7 V and 20 nA, respectively. In contrast, no initial filament was formed under the same electrical conditions in vacuum, even after multiple attempts. Further increases in the applied voltage resulted in a very low LRS, and a subsequent reset attempt required high reset voltages that led to a permanent breakdown.

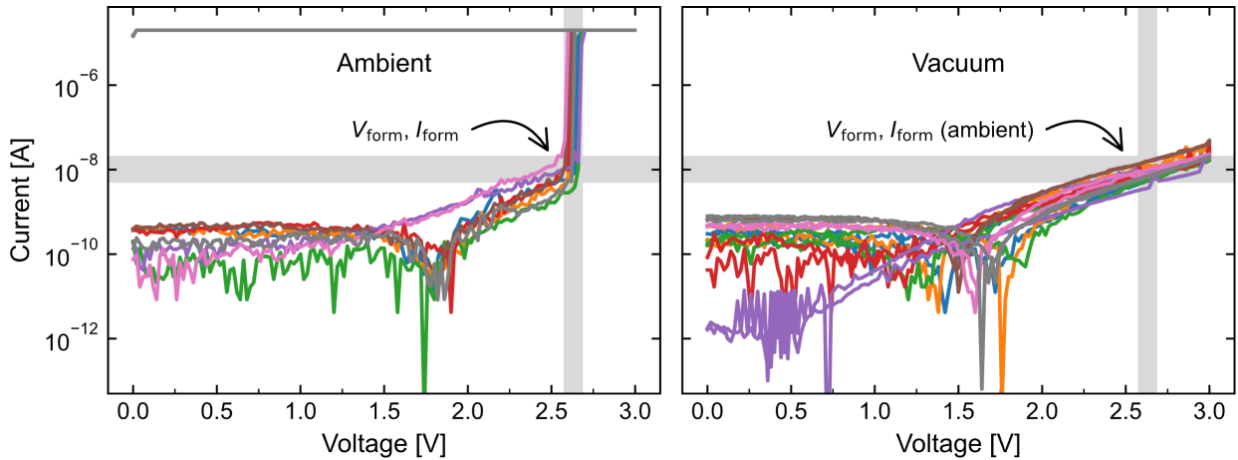

**Figure 3:** Forming attempts of eight devices each under ambient (left) and vacuum (right) conditions. The gray areas indicate the forming voltage and forming current ranges that lead to successful forming under ambient conditions. The parameters used under ambient conditions did not lead to forming under vacuum.

A set kinetics plot describes the relationship between the set time and the applied voltage of a memristor [20]. In ECM cells, this relationship allows insights into the time-limiting processes during the resistive switching event. These processes can be nucleation-limited, redox-based, electron transfer-limited, and ion migration-limited [21]. While the change in the applied voltage is small, the set time varies by orders of magnitude. We applied voltage pulses with pulse widths ranging from minutes down to the nanosecond timescale. Fig. 4a shows an example of a set pulse with a pulse amplitude of 5 V, leading to a switching time $t_{set}$ = 6.4 μs. The current limit was set to 50 μA during the measurement. The current increases abruptly due to the set transition after 6.4 μs, leading to a current overshoot, but is limited to 50 μA after ~ 900 ns by the current-limiting circuit. Repeating this experiment for different pulse heights results in a characteristic set kinetics plot (Fig. 4b). The relationship between the set time and the applied voltage is well described by a simplification of the Butler-Volmer equation [22,23], as implemented in the JART ECM model [16,17] and is illustrated by the orange line in Fig. 4b. The model assumes three different contributions to the ionic device current during resistive switching: (1) the current flow from the electron transfer reactions (redox reactions) at the active electrode/h-BN interface ($I_{ac}$); (2) the current flow at the h-BN/filament tip interface ($I_{fil}$); and (3) the current flow caused by the ion migration process ($I_{hop}$). The redox reaction current is described by the Butler-Volmer equation, highlighting the importance of redox reactions during resistive switching:

$$I_{ac/fil} = \pm zeck_{0,et}A_{ac/fil}\exp\left(-\frac{\Delta G_{et}}{k_BT}\right) \cdot \left[\exp\left(\frac{(1-\alpha_{et})ze}{k_BT}\eta_{ac/fil}\right) - \exp\left(-\frac{\alpha_{et}ze}{k_BT}\eta_{ac/fil}\right)\right]. \quad (1)$$

Here, $z$ is the charge transfer number, $e$ is the elementary charge, $c$ is the nickel ion concentration, $k_{0,et}$ is the electron transfer reaction rate, $A_{ac/fil}$ are the (circular) active electrode reaction area and the filament area, respectively, $\Delta G_{et}$ is the electron transfer activation barrier, $k_B$ is the Boltzmann constant, $T$ is the temperature, $\alpha_{et}$ is the electron transfer coefficient, and $\eta_{ac/fil}$ is the respective electron transfer overpotential. The current flow caused by the ion migration process is described by the Mott-Gurney law [24]:

$$I_{hop} = 2zecaf\exp\left(-\frac{\Delta G_{hop}}{k_BT}\right)A_{is}\sinh\left(\frac{aze}{2k_BT}\frac{\eta_{hop}}{x}\right), \quad (2)$$

where $a$ is the mean ion hopping distance, $f$ is the hopping attempt frequency, $\Delta G_{hop}$ is the ion hopping activation barrier, $A_{is}$ is the cross-sectional area of the ion migration in the switching layer, $\eta_{hop}$ is the ion hopping overpotential and $x$ is the tunneling gap between the tip of the filament and the active electrode. By inserting one of the ionic currents $I_{ac}$, $I_{fil}$ or $I_{hop}$ as the current density $j_{ion}$ into equation 3,

$$\frac{\partial x}{\partial t} = -\frac{M_{Me}}{ze\rho_{m,Me}} j_{ion}, \qquad (3)$$

the transient evolution of the tunneling gap $x$ depending on the ionic current density $j_{ion}$ is described, and thus, the evolution of the filament. In equation 3, $M_{Me}$ and $\rho_{m,Me}$ are the molecular mass and mass density of nickel, respectively. The switching time $t_{set}$ is mainly determined by equation (3). All the model parameters are summarized in Table 1, and the full model is presented in[16,17].

**Table 1:** Model parameters, descriptions and values.

| Symbol/Parameter | Value | Symbol/Parameter | Value |
|---|---|---|---|
| $M_{me}$: molecular mass of Nickel | $9.75 \cdot 10^{-26}$ kg | $r_{ac}$: radius of active electrode reaction area $A_{ac}$ | 10 nm |
| $z$: charge transfer number | 2 | $r_{fil}$: filament radius of a filament area $A_{fil}$ | 1 nm |
| $\rho_{m,me}$: mass density of Nickel | $8.91 \cdot 10^3$ kg/m$^3$ | $r_{is}$: cross-sectional radius of ion hopping process | 10 nm |
| $m_r$: relative electron mass | 2.21 | $L$: length of switching/insulating layer | 3 nm |
| $\Delta W_0$: eff. tunneling barrier height | 3.71 eV | $\rho_{fil}$: filament's electronic resistivity | $6.85 \cdot 10^{-8}$ Ωm |
| $\alpha_{et}$: electron transfer coefficient | 0.07 | $R_{el}$: resistance of electrodes | 40 Ω |
| $k_{0,et}$: electron transfer reaction rate | $1 \cdot 10^3$ m/s | $R_S$: series resistance | 120 Ω |
| $c$: Nickel ion concentration | $1 \cdot 10^{24}$ 1/m$^3$ | $T$: temperature | 298 K |
| $\Delta G_{et}$: electron transfer activation barrier | 0.65 eV | $t_{0,nuc}$: prefactor of nucleation time | $6 \cdot 10^{-10}$ s |
| | | $\Delta G_{nuc}$: nucleation activation energy | 0.8 eV |
| $m_e$: electron mass | $9.11 \cdot 10^{-31}$ kg | $N_c$: number of atoms for stable nucleus | 3 |
| $h$: Planck constant | $6.63 \cdot 10^{-34}$ J s | $\alpha_{nuc}$: electron transfer coefficient for nucleation | 0.3 |
| $e$: elementary charge | $1.60 \cdot 10^{-19}$ C | | |
| $k_B$: Boltzmann constant | $1.38 \cdot 10^{-23}$ J/K | $a$: mean ion hopping distance | 0.17 nm |
| $\eta_{ac/fil/hop}$: respective electron transfer overpotential | | $f$: ion hopping attempt frequency | $1 \cdot 10^{13}$ Hz |
| | | $\Delta G_{hop}$: ion migration barrier | 0.22 eV |

We fitted the experimental data by applying equation 3 with two different Ni-ion concentrations $c$, as shown in Fig. 4b. The first slope of the simulated curves is determined by the redox-based electron transfer processes. Here, this is the first regime for small or intermediate applied voltages, and the redox processes are the slowest of the ionic processes. The second slope represents a switching time limitation through a combination of redox processes at the interfaces and ionic migration processes through the switching layer. Both ionic processes simultaneously limit the switching speed for high voltages [16,17]. Valov and Tsuruoka extensively studied the electrochemical processes in $SiO_2$- and $Ta_2O_5$-based ECM cells that two electrochemical processes are relevant for filament formation during the initial forming of a pristine device and after strong reset events [25]: the anodic oxidation of the active metal electrode at the active electrode/insulator interface, here $Ni \rightarrow Ni^{2+} + 2e^-$, and the reduction reaction at the insulator/inert electrode interface, including water $2H_2O + 2e^- \rightleftharpoons 2OH^- + H_2$. This counterelectrode half-cell reaction is essential to sustain the redox reaction until Ni ions migrate through the insulating layer and reach the counter electrode for reduction. Once they do, Ni ions are reduced, and the processes of Ni and water reduction occur simultaneously. Therefore, the reaction involving water is crucial as an initial counterpart reaction to maintain charge neutrality within the cell. In a vacuum, there is less humidity. Thus, this water reduction reaction and, consequently, the forming and filament growth are suppressed. Furthermore, the concentration of Ni ions is very limited due to the absence of charge compensation mechanisms [20].

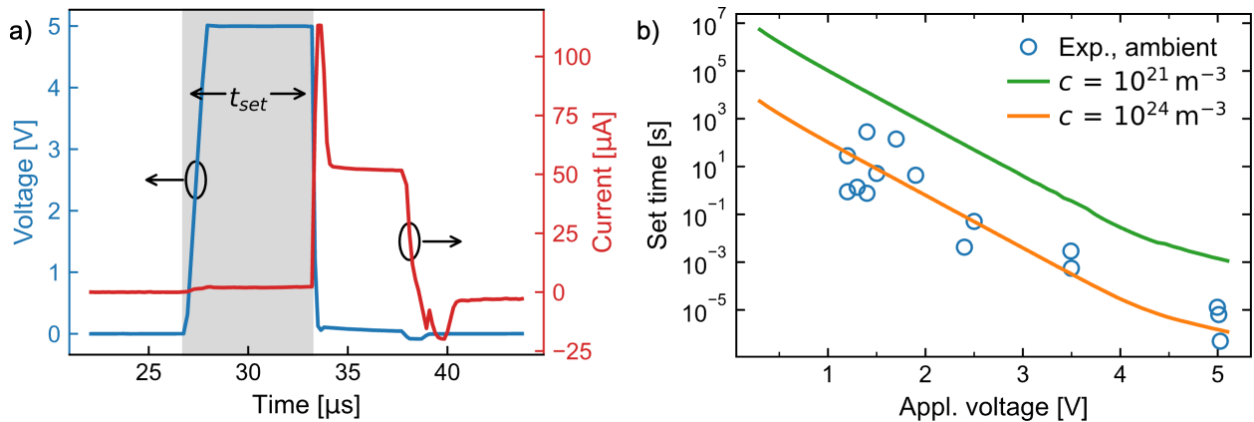

**Figure 4:** a) Example set pulse and current response with a pulse width of 10 µs. The voltage pulse appears to be shorter because the current limiting circuit decreases the applied voltage after the set event. The current limit was set to 50 µA. The gray area indicates the set time $t_{set}$. b) Set kinetics

plot with extracted set times $t_{set}$ versus the applied voltage pulse height. The trend of the experimental data under ambient conditions can be described by the Butler-Volmer equation (orange line). A decreased Ni ion concentration $c$ reflects vacuum-like conditions, leading to significantly longer switching times (green line).

In the simulation (Fig. 4b), the effect of vacuum, i.e., less humidity, is considered with a lower Ni ion concentration $c$ ($10^{21}$ m$^{-3}$, green) than that of the ambient atmosphere ($c = 10^{24}$ m$^{-3}$, orange). The simulation results reveal that the switching time drastically increases with decreasing Ni ion concentration. Only very high voltages enable switching, and a long switching time $t_{set}$ is needed. These simulations support our argument that a lack of water in the environment of our cells limits the formation of a Ni filament and can explain the suppressed forming of our cells in a vacuum (Fig. 3).

In conclusion, memristors based on multilayer h-BN sandwiched between Pd and Ni electrodes were investigated. The memristors exhibit repeatable bipolar resistive switching under voltage stress applied to the Ni electrode. Comparative electrical *I-V* measurements under ambient and vacuum conditions reveal the necessity of ambient air molecules for the resistive switching of the cells. The set kinetics of the memristors were investigated by switching the devices with voltage pulses ranging from minutes to nanoseconds. They can be described by the Butler-Volmer equation, which reflects the occurrence of redox reactions during resistive switching. Simulations confirm that the absence of humidity leads to drastically longer switching times and thus can explain the suppressed forming of our cells in vacuum. These findings have important implications for potential applications of h-BN memristors, as computer chips in end-user applications are usually encapsulated to protect them from atmospheric influences. Further experiments are needed to analyze the exact reaction pattern during switching. Knowledge of the reaction equations may allow the material stack to be adapted so that resistive switching is still possible without water molecules.

## Methods

### Device Fabrication

2 cm × 2 cm silicon (Si) samples with 300 nm thermally oxidized silicon dioxide ($SiO_2$) were used as the starting substrates. The bottom electrodes (BEs) were defined via optical lithography, followed by 10 nm titanium (Ti) and 40 nm palladium (Pd) *in-situ* deposition via electron beam evaporation. A subsequent lift-off process defined the contact areas. h-BN was grown via metalorganic chemical vapor deposition (MOCVD) on c-plane sapphire (with a 0.2° offcut toward the m-plane) in an AIXTRON Close Coupled Showerhead (CCS®) 2D R&D reactor in a 19x2" configuration. A continuous h-BN film was formed via a one-step growth process at 1350°C and 500 mbar pressure for 1600 s using borazine ($B_3N_3H_6$) as a precursor material and $H_2$ as a carrier gas. The film was subsequently transferred to the sample via a wet transfer approach. The top electrodes (TEs) were defined via a second optical lithography step, direct current (DC) sputtering of 30 nm nickel (Ni) and 30 nm aluminum (Al, *in situ*), and a subsequent lift-off process.

AUTHOR INFORMATION

**Corresponding Author**

*Max C. Lemme[a, d]

[a] Chair of Electronic Devices, RWTH Aachen University, Otto-Blumenthal-Str. 25, 52074 Aachen, Germany.

[d] AMO GmbH, Advanced Microelectronic Center Aachen, Otto-Blumenthal-Str. 25, 52074 Aachen, Germany.

Email: max.lemme@eld.rwth-aachen.de

**Author Contributions**

The experiments were conceived by L.V., R.W.A., S.M., M.H. and M.C.L. The fabrication of the samples was carried out by L.V. h-BN was synthesized by S.K. and S.P. The electrical measurements were performed by L.V. Raman spectra and AFM images were obtained by L.V. The simulations were conducted by R.W.A., A.B., and S.M. The data were analyzed by L.V., D.B., S.C., J.L., R.W.A., A.B., S.M. and M.C.L. All authors collaborated on the interpretation of the experiments. The manuscript was written and revised by all. The work was supervised by S.M. and M.C.L.

**Competing Interests**

M.C.L. is the managing director of the non-profit company AMO GmbH, whereas S.P., S.K., and M.H. are employees of AIXTRON SE. Both companies are partners of several publicly funded research projects and are working on the growth and device integration of 2D materials. L.V., R.W.A., A.B., D.B., S.C., J.L., and S.M. have no competing interests.

## ACKNOWLEDGMENT

We gratefully acknowledge financial support from the German Federal Ministry of Education and Research (BMBF) within the projects NEUROTEC 2 (16ME0399, 16ME0398K, 16ME0400) and NeuroSys (03ZU1106AA, 03ZU1106BA, 03ZU2106AE, 03ZU2106AA), from the European Union's Horizon Europe research and innovation program under the project ENERGIZE (101194458), and Deutsche Forschungsgemeinschaft (DFG) within the project SPP2262 MEMMEA, project no. 441918103 (MemrisTec).

## ABBREVIATIONS

2D, two-dimensional; h-BN, hexagonal boron nitride; ECM, electrochemical metallization; Si, silicon; $SiO_2$, silicon dioxide; BE, bottom electrode; Ti, titanium; Pd, palladium; MOCVD, metalorganic chemical vapor deposition; CCS®, Close Coupled Showerhead; $B_3N_3H_6$, borazine; TE, top electrode; DC, direct current; Ni, nickel; Al, aluminum; AFM, atomic force microscopy; HRS, high resistance state; *I-V*, current-voltage; LRS, low resistance state